\documentclass[12pt,a4paper]{article}
\usepackage{fullpage}
\usepackage{amsbsy}
\usepackage{amsfonts}
\usepackage{amssymb}
\usepackage{url}
\usepackage[square,comma,numbers]{natbib}
\begin{document}

\begin{flushright}
\textsf{6 February 2003}
\\
\textsf{hep-ph/0302045}
\end{flushright}

\vspace{1cm}

\begin{center}
\large
\textbf{Coherence in Neutrino Interactions}
\normalsize
\\[0.5cm]
\large
C. Giunti
\normalsize
\\[0.5cm]
INFN, Sezione di Torino, and Dipartimento di Fisica Teorica,
\\
Universit\`a di Torino,
Via P. Giuria 1, I--10125 Torino, Italy
\\[0.5cm]
\begin{minipage}[t]{0.8\textwidth}
\begin{center}
\textbf{Abstract}
\end{center}
The claim in hep-ph/0301231 is refuted in a pedagogical way.
It is explicitly shown that
extremely relativistic neutrinos produced in pion decay
are correctly described by the standard
flavor neutrino states
which are a coherent superposition of massive neutrino states.
\end{minipage}
\end{center}

The author of Ref.~\cite{Field:2003tt} wrote:
\begin{quote}
If neutrinos are massive,
`lepton flavor eigenstates'
are absent from the amplitudes of all
Standard Model processes.
\end{quote}
His argument follows from a faulty calculation of the amplitude in
the pion decay process
\begin{equation}
\pi^- \to \ell^- + \bar\nu_\ell
\,,
\label{001}
\end{equation}
where $\ell=e,\mu$.

According to the standard theory of neutrino mixing
\cite{Eliezer:1976ja,Fritzsch:1976rz,Bilenky:1976yj,Bilenky:1978nj},
the antineutrino produced in the process (\ref{001})
is described by the
\emph{coherent superposition of massive antineutrino states}
\begin{equation}
|\bar\nu_\ell\rangle
=
\sum_k U_{\ell k} |\bar\nu_k\rangle
\,,
\label{002}
\end{equation}
where $U$ is the unitary mixing matrix.
Sometimes such a state
is called
``flavor state'',
or
``flavor eigenstate'',
or
``lepton flavor eigenstate''.
The author of Ref.~\cite{Field:2003tt}
claimed that
\begin{quote}
the introduction of such
`lepton flavor eigenstates'
as linear superpositions of neutrino mass eigenstates
leads to predictions that are excluded by experiment.
\end{quote}

In order to confute the argument presented in Ref.~\cite{Field:2003tt},
let us calculate in a correct way the amplitude
of the pion decay process (\ref{001})
in the case of neutrino mixing.

At the first order of perturbation theory,
the amplitude of the pion decay process (\ref{001}) is
\begin{equation}
\mathcal{A}_{\pi^- \to \ell^- \bar\nu_\ell}
=
\langle \ell^-, \bar\nu_\ell | -i \int \mathrm{d}^4x \, \mathcal{H}_{\mathrm{I}}^{\mathrm{CC}}(x) | \pi^- \rangle
\,,
\label{003}
\end{equation}
with the Charged-Current Standard Model weak interaction Hamiltonian
\begin{eqnarray}
\mathcal{H}_{\mathrm{I}}^{\mathrm{CC}}(x)
& = &
\frac{G_F}{\sqrt{2}}
\sum_{\ell=e,\mu,\tau}
\bar\ell(x)
\,
\gamma^{\rho}
\left( 1 - \gamma_5 \right)
\nu_{\ell}(x)
\,
J_{\rho}(x)
+
\mathrm{h.c.}
\nonumber
\\
& = &
\frac{G_F}{\sqrt{2}}
\sum_{\ell=e,\mu,\tau}
\sum_k
\bar\ell(x)
\,
\gamma^{\rho}
\left( 1 - \gamma_5 \right)
U_{{\ell}k}
\,
\nu_k(x)
\,
J_{\rho}(x)
+
\mathrm{h.c.}
\,,
\label{004}
\end{eqnarray}
where
$G_F$ is the Fermi constant
and
$J_{\rho}(x)$
is the hadronic weak charged current.
Notice that we have expressed the Hamiltonian in terms of the
neutrino fields $\nu_k(x)$ with mass $m_k$,
that create the corresponding massive antineutrino states
$|\bar\nu_k\rangle$
in Eq.~(\ref{002}).

Using
Eqs.~(\ref{002}) and (\ref{004}),
the amplitude (\ref{003})
can be written as
\begin{equation}
\mathcal{A}_{\pi^- \to \ell^- \bar\nu_\ell}
=
- i \frac{G_F}{\sqrt{2}}
\int \mathrm{d}^4x
\sum_{k,j} U_{\ell k}^* U_{\ell j}
\langle \ell^-, \bar\nu_k |
\bar\ell(x)
\,
\gamma^{\rho}
\left( 1 - \gamma_5 \right)
\nu_j(x)
\,
J_{\rho}(x)
| \pi^- \rangle
\,.
\label{005}
\end{equation}
Notice the presence in Eq.~(\ref{005})
of two elements of the mixing matrix,
one coming from the mixing of the states in Eq.~(\ref{002})
and the other coming from the mixing of the fields in Eq.~(\ref{004}).
Since only the state
$|\bar\nu_k\rangle$
is a quantum of the corresponding massive neutrino field
$\nu_k(x)$,
the matrix element in Eq.~(\ref{005})
is proportional to $\delta_{kj}$,
\begin{equation}
\langle \ell^-, \bar\nu_k |
\bar\ell(x)
\,
\gamma^{\rho}
\left( 1 - \gamma_5 \right)
\nu_j(x)
\,
J_{\rho}(x)
| \pi^- \rangle
\propto
\delta_{kj}
\,,
\label{005a}
\end{equation}
leading to
\begin{equation}
\mathcal{A}_{\pi^- \to \ell^- \bar\nu_\ell}
=
- i \frac{G_F}{\sqrt{2}}
\int \mathrm{d}^4x
\sum_{k}|U_{\ell k}|^2
\langle \ell^-, \bar\nu_k |
\bar\ell(x)
\,
\gamma^{\rho}
\left( 1 - \gamma_5 \right)
\nu_k(x)
\,
J_{\rho}(x)
| \pi^- \rangle
\,.
\label{005b}
\end{equation}

Since the neutrino masses are much smaller than the energy released
in the pion decay process (\ref{001}),
the dependence of the matrix elements
in Eq.~(\ref{005b})
on the corresponding neutrino mass $m_k$
can be neglected
and the matrix elements
can be approximated with the matrix element in the case of massless neutrinos,
\begin{equation}
\langle \ell^-, \bar\nu_k |
\bar\ell(x)
\,
\gamma^{\rho}
\left( 1 - \gamma_5 \right)
\nu_k(x)
\,
J_{\rho}(x)
| \pi^- \rangle
\simeq
\langle \ell^-, \bar\nu |
\bar\ell(x)
\,
\gamma^{\rho}
\left( 1 - \gamma_5 \right)
\nu(x)
\,
J_{\rho}(x)
| \pi^- \rangle
\,,
\label{005c}
\end{equation}
where
$\nu(x)$
is a massless neutrino field with antineutrino quanta $|\bar\nu\rangle$.
In this case,
the matrix elements
in Eq.~(\ref{005b})
can be extracted from the sum over the mass index $k$
and,
using the unitarity relation
$ \sum_k |U_{\ell k}|^2 = 1 $,
one obtains
\begin{equation}
\mathcal{A}_{\pi^- \to \ell^- \bar\nu_\ell}
\simeq
- i \frac{G_F}{\sqrt{2}}
\int \mathrm{d}^4x
\langle \ell^-, \bar\nu |
\bar\ell(x)
\,
\gamma^{\rho}
\left( 1 - \gamma_5 \right)
\nu(x)
\,
J_{\rho}(x)
| \pi^- \rangle
\,,
\label{006}
\end{equation}
which is the amplitude of the pion decay process (\ref{001})
in the case of massless neutrinos.

Therefore,
in the realistic case of extremely relativistic neutrinos
the pion decay rate calculated
in a correct way
using the flavor antineutrino state (\ref{002})
practically coincides with the standard pion decay rate calculated
assuming massless neutrinos.
It is pretty obvious that a similar conclusion
holds for all weak processes.

Somehow the author of Ref.~\cite{Field:2003tt}
missed the quadratic dependence of the amplitude on the elements of the mixing matrix
and obtained an absurd result:
if $|U_{\ell k}|^2$
in Eq.~(\ref{005b})
were replaced by $U_{\ell k}$,
as sadly happens in Ref.~\cite{Field:2003tt},
the amplitude would depend on the elements of the mixing matrix
even in the case of massless neutrinos.
This is clearly nonsense,
because in the case of massless neutrinos
there is no mixing.
Nevertheless, the author of Ref.~\cite{Field:2003tt}
considered it seriously
and confronted the resulting decay rate with experimental data.
Since the measured ratio of the
$\pi^- \to e^- + \bar\nu_e$
and
$\pi^- \to \mu^- + \bar\nu_\mu$
decay rates is incompatible with the wrong ratio
calculated in Ref.~\cite{Field:2003tt},
the author of Ref.~\cite{Field:2003tt}
erroneously
claimed that the neutrino produced in the process (\ref{001})
is not described by the coherent superposition of massive antineutrino states
in Eq.~(\ref{002}).

Having clarified the main mistake in Ref.~\cite{Field:2003tt},
some further remarks are in order:

\begin{enumerate}
\item \label{coherence}
The author of Ref.~\cite{Field:2003tt}
denies the coherent character of
massive neutrino states produced in weak processes,
but somewhat manages to get neutrino oscillations
(albeit with a wrong phase, see item~\ref{phase} below).
Obviously there is a contradiction:
since neutrino oscillations are due to the interference of
different massive neutrinos their coherence is required.
\item \label{states}
Ref.~\cite{Giunti:1992cb}
was cited in an improper way in Ref.~\cite{Field:2003tt},
in connection with the sentence
``the unphysical nature of coherent states of neutrinos
of different mass was also discussed in the literature''.
On the contrary,
in Ref.~\cite{Giunti:1992cb} it is explicitly written that
neutrinos produced in weak interaction processes
are described by a (coherent) superposition of massive neutrino states,
which in the realistic limit of extremely relativistic neutrinos reduces
to the standard expression (\ref{002}).
These flavor neutrino states (called ``weak-process states'' in Ref.~\cite{Giunti:1992cb})
were recently calculated in Ref.~\cite{Giunti:2002xg}
in a quantum field theoretical wave packet approach.

In Ref.~\cite{Giunti:1992cb}
it has been shown that the flavor state $|\bar\nu_\ell\rangle$
in Eq.~ (\ref{002})
is not a quantum of the field $\nu_\ell(x)$,
and
the field $\nu_\ell(x)$
is not quantizable\footnote{
In Ref.~\cite{Giunti:1992cb}
it has been implicitly assumed that a flavor neutrino state
is a superposition of massive neutrino states,
excluding any antineutrino component.
If this assumption is relaxed,
the flavor fields $\nu_\ell(x)$
can be quantized,
as shown in Ref.~\cite{Blasone:1995zc}.
However,
the physical meaning of a superposition of massive neutrino
and antineutrino states
is unclear to us.
}
because it does not have a definite mass
(different flavor neutrino fields are coupled through the non-diagonal mass term in
the Lagrangian).
Obviously, this does not mean that a flavor state $|\bar\nu_\ell\rangle$
defined as a coherent superposition
of massive antineutrino states
is ``unphysical'', as claimed in Ref.~\cite{Field:2003tt}.
\item \label{phase}
The final goal of Ref.\cite{Field:2003tt}
is to renew the claim of a factor of two mistake in the standard phase
of neutrino oscillations.
This claim has been already confuted in
Refs.~\cite{Giunti:2000kw,Giunti:2002ee,hep-ph/0211241}.
\end{enumerate}

In conclusion,
we would like to express a note of praise for the electronic archives,
which allow a wide diffusion of all kind of ideas
that stimulate interesting thinking.

\end{document}